\documentclass[fleqn,usenatbib]{mnras}

\usepackage{newtxtext,newtxmath}

\usepackage[T1]{fontenc}

\DeclareRobustCommand{\VAN}[3]{#2}
\let\VANthebibliography\thebibliography
\def\thebibliography{\DeclareRobustCommand{\VAN}[3]{##3}\VANthebibliography}

\usepackage{graphicx}	
\usepackage{amsmath}	
\newcommand\target{{MAXI~J1348$-$630}}
\newcommand\nicer{{\it NICER}}
\defcitealias{Zhang2020a}{Paper-I}

\title[Type-A QPO in \target]{Type-A quasi-periodic oscillation in the black hole transient \target}

\author[Zhang et al.]{
Liang Zhang$^{1},$\thanks{E-mail: zhangliang@ihep.ac.cn}
Mariano M\'{e}ndez,$^{2}$
Federico Garc\'ia,$^{3}$
Yuexin Zhang,$^{2}$
Ruican Ma,$^{1}$
\newauthor
Diego Altamirano,$^{4}$
Zi-Xu Yang,$^{1}$
Xiang Ma,$^{1}$
Lian Tao,$^{1}$
Yue Huang,$^{1}$
Shumei Jia,$^{1}$
\newauthor
Shuang-Nan Zhang,$^{1}$
Jinlu Qu,$^{1}$
Liming Song$^{1}$
and
Shu Zhang$^{1}$
\\
$^{1}$Key Laboratory of Particle Astrophysics, Institute of High Energy Physics, Chinese Academy of Sciences, Beijing 100049, China\\
$^{2}$Kapteyn Astronomical Institute, University of Groningen, Postbus 800, 9700 AV Groningen, The Netherlands\\
$^{3}$Instituto Argentino de Radioastronom\'{i}a (CCT La Plata, CONICET; CICPBA; UNLP), C.C.5, (1894) Villa Elisa, Buenos Aires, Argentina\\
$^{4}$Physics and Astronomy, University of Southampton, Southampton, Hampshire SO17 1BJ, UK \\
}

\date{Accepted XXX. Received YYY; in original form ZZZ}

\pubyear{2023}

\begin{document}
\label{firstpage}
\pagerange{\pageref{firstpage}--\pageref{lastpage}}
\maketitle

\begin{abstract}
We present a detailed analysis of the spectral and timing characteristics of a 7-Hz type-A quasi-periodic oscillation (QPO) detected in \nicer\ observations of the black hole X-ray binary \target\ during its high-soft state. The QPO is broad and weak, with an integrated fractional rms amplitude of 0.9 per cent in the 0.5--10 keV band. Thanks to the large effective area of \nicer, combined with the high flux of the source and a relatively long accumulative exposure time, we construct the first rms and phase-lag spectra for a type-A QPO. Our analysis reveals that the fractional rms amplitude of the QPO increases with energy from below 1 per cent at 1 keV to $\sim$3 per cent at 6 keV. The shape of the QPO spectrum is similar to that of the Comptonised component, suggesting that the Comptonised region is driving the variability. The phase lags at the QPO frequency are always soft taking the lowest energy as reference. By jointly fitting the time-averaged spectrum of the source and the rms and phase-lag spectra of the QPO with the time-dependent Comptonisation model \textsc{vkompthdk}, we find that the radiative properties of the type-A QPO can be explained by a vertically extended Comptonised region with a size of $\sim$2300 km.   
\end{abstract}




\section{Introduction}

Black hole X-ray binaries (BHXBs) are unique laboratories to study accretion onto compact objects and to test General Relativity in the strong-field regime. 
These systems are typically transient sources that transition between different accretion states with well-defined spectral-timing properties during an outburst \citep[see][for a recent review]{Belloni2016}. 
Following the classification of \citet{Homan2005}, four main states can be identified: the low-hard state (LHS), the hard-intermediate state (HIMS), the soft-intermediate state (SIMS), and the high-soft state (HSS).
Throughout most outbursts, BHXBs evolve from the LHS, through the HIMS and SIMS, eventually reaching the HSS during the outburst rise. During the outburst decay, the sources generally return to the LHS in reverse order \citep[e.g.,][]{Dunn2010, Motta2011, Munoz2011, Zhang2020a}.

BHXBs typically display strong variability on sub-second timescales. Notably, these systems often show low-frequency QPOs in the power density spectrum (PDS) \citep[see][for a recent review]{Ingram2019}. These QPOs have been observed in nearly all BHXBs and their characteristics are highly correlated with spectral states \citep{Belloni2005, Motta2011, Munoz2011, Zhang2020a}.
Based on differences in the shape of the PDS, low-frequency QPOs in BHXBs can be classified into three types: A, B, and C \citep{Wijnands1999, Remillard2002, Casella2005}.
Type-C QPOs usually appear in the LHS and HIMS, and are characterized by a high-amplitude (up to 20 per cent rms in the full RXTE/PCA band), narrow peak ($Q$\footnote{$Q=\nu_{0}/{\rm FWHM}$, where $\nu_{0}$ is the centroid frequency of the QPO and FWHM is its full width at half maximum.}$\gtrsim$8) with a variable frequency ranging from 0.1 to 30 Hz, superposed on a strong band-limited noise continuum. It has been suggested that type-C QPOs may arise from the Lense-Thirring precession of the hot inner flow or jet \citep{Ingram2009, You2018, Ma2021}, although this model faces theoretical and observational challenges \citep{Marcel2021, Nathan2022}. 
Type-B QPOs are observed in the short-lived SIMS. They appear in the PDS as a relatively high-amplitude (up to 5 per cent rms in the full RXTE/PCA band) and narrow peak ($Q\gtrsim6$) at around 4--6 Hz, accompanied by a weak broadband noise component. Type-B QPOs are thought to be associated with the production of jet outflows as they are found to occur close in time to the launch of discrete jet ejecta \citep{Homan2020, Russell2020, Carotenuto2021}.
Compared to type-C and type-B QPOs, type-A QPOs are the least frequently observed and have been rarely studied  \citep[e.g.,][]{Homan2001, Remillard2002, Casella2004, Motta2011, Zhang2020a}. They are normally detected in the HSS, where the energy spectrum is dominated by thermal emission from the accretion disc. Type-A QPOs appear as a weak (fractional rms of $\lesssim$2 per cent in the full RXTE/PCA band) and broad peak ($Q\lesssim3$) with a centroid frequency of $\sim$6--8 Hz, coincident with a very weak broadband noise component. To date, no comprehensive model has been proposed for type-A QPOs.

Examining the energy-dependent properties of QPOs, such as the fractional rms amplitude and phase-lag spectra, can offer valuable insights into the physical origin of QPOs, as well as the geometry of the region where they originate \citep[e.g.,][]{Reig2000, Rodriguez2004, Eijnden2017, Zhang2017, Huang2018, Jithesh2021, Ma2023}. 
Recent findings from {\it Insight}-HXMT have revealed that, for both type-C and type-B QPOs, the fractional rms amplitudes exhibit an initial increase with energy, followed by a plateau up to energies above 100 keV \citep{Huang2018, Ma2021, Liu2022}. At these higher energies, the fractional rms amplitudes typically reach 10--20 per cent. These results suggest that the QPOs are generated in the Comptonised region, either the corona or the base of the jet, rather than the accretion disc \citep{Sobolewska2006, Kylafis2020, Ma2021, Mendez2022}.
The shape of the phase-lag spectrum of low-frequency QPOs is known to be more complex \citep[e.g.,][]{Wijnands1999, Lin2000, Eijnden2017}. For type-C QPOs, the shape of the lag spectrum undergoes significant evolution during the transition from the LHS to the HIMS \citep{Zhang2017, Ma2023}. On the other hand, type-B QPOs usually show a "U"-shaped lag spectrum \citep{2018ApJ...865L..15S, Belloni2020, Peirano2023}.
To date, there is no report of an rms or lag spectrum for a type-A QPO.
Based on the work by \citet{Karpouzas2020}, recently, \citet{Bellavita2022} developed a time-dependent Comptonisation model that explains the energy-dependent properties of the low-frequency QPOs in BHXBs. In this model, the radiative properties of the variability are due to coupled oscillations between the corona and the disc \citep{Mastichiadis2022}. 
This model has been applied successfully to the type-C and type-B QPOs observed in a number of BHXBs \citep{Garcia2021, Garcia2022, Mendez2022, Zhang2022, Ma2023b, Peirano2023, Zhang2023}.

\target\ is a Galactic BHXB discovered with MAXI/GSC on 2019 January 26 \citep{Yatabe2019, Tominaga2020}, when the source entered into a bright outburst with a peak flux of approximately 4 Crab (2--20 keV). 
\target\ is located at a relatively close distance, between 2.2 and 3.4 kpc \citep{Chauhan2021, Lamer2021}. As of now, no dynamical constraint on the mass of the compact object has been obtained.
A comprehensive spectral-timing analysis conducted with \nicer\ revealed that \target\ underwent an initial full outburst, displaying all the canonical spectral states of BHXBs. Following the full outburst, the source re-brightened a few times, during which it remained in the LHS \citep[hereafter \citetalias{Zhang2020a}]{Zhang2020a}. 
Different types of low-frequency QPOs have been observed at different phases of the outburst (see \citetalias{Zhang2020a}).
\citet{Alabarta2022} studied the energy-dependent properties of the type-C QPOs in \target\ using \nicer\ data. They found that both the energy-dependent fractional rms amplitude and the phase lags of the type-C QPO can be explained within the framework of Comptonisation. 
Additionally, a strong type-B QPOs at $\sim$4.5 Hz was observed in the SIMS of \target\ \citep{Belloni2020, Garcia2021, Zhang2021, Liu2022}, exhibiting rapid appearance/disappearance on short timescales of a few tens of seconds \citep{Zhang2021, Liu2022}.
\citet{Belloni2020} calculated the energy-dependent fractional rms and phase lags of the type-B QPO using \nicer\ data. They found that the fractional rms amplitude increases with energy and the phase lags at the QPO frequency are positive with respect to the reference band 2--3 keV.
\citet{Garcia2021} interpreted the energy-dependent properties of the type-B QPO using a two-component Comptonisation model.
\citet{Zhang2021} found that the transient nature of the type-B QPO is associated with a redistribution of accretion power between the disc and Comptonised emission \citep[see also][]{Yang2023}.

In \citetalias{Zhang2020a}, we also reported the detection of a 7-Hz type-A QPO in multiple \nicer\ observations of \target. In this work, we conduct a comprehensive spectral-timing analysis to investigate the properties and characteristics of the type-A QPO in \target.

\section{Observations and data reduction}

\nicer\ is a soft X-ray telescope onboard the International Space Station (ISS) \citep{Gendreau2016}. It provides high throughput in the 0.2--12 keV energy band with an absolute timing precision of $\sim$100 ns, making it an ideal instrument for studying fast X-ray variability. 
The X-ray Timing Instrument (XTI) of \nicer\ comprises an array of 56 co-aligned concentrator X-ray optics. Each optic is paired with a single-pixel silicon drift detector. Presently, 52 detectors are operational with a peak effective area of $\sim$1900 $\rm{cm}^2$ at 1.5 keV.

In this study, we present the analysis of \nicer\ observations during the soft state of \target\ to search for type-A QPOs. These data correspond to ObsIDs 1200530126--1200530128 and 2200530101--2200530126. To prevent telemetry saturation and accommodate the high source flux, some of the detectors were switched off during this period. The data were reprocessed using the \nicer\ software tools NICERDAS version 2022-12-16\_V010a, along with the calibration database (CALDB) version xti20221001. We filtered the data with the standard screening criteria via the \texttt{nicerl2} task. Additionally, we filtered out times of high background by removing all intervals with a 13--15 keV rate (where no source contribution is expected) exceeding 2 cts s$^{-1}$. The good time intervals of each observation were further divided into multiple continuous data segments based on the orbit of the ISS.

\begin{figure}
\begin{center}
\resizebox{1\columnwidth}{!}{\rotatebox{0}{\includegraphics{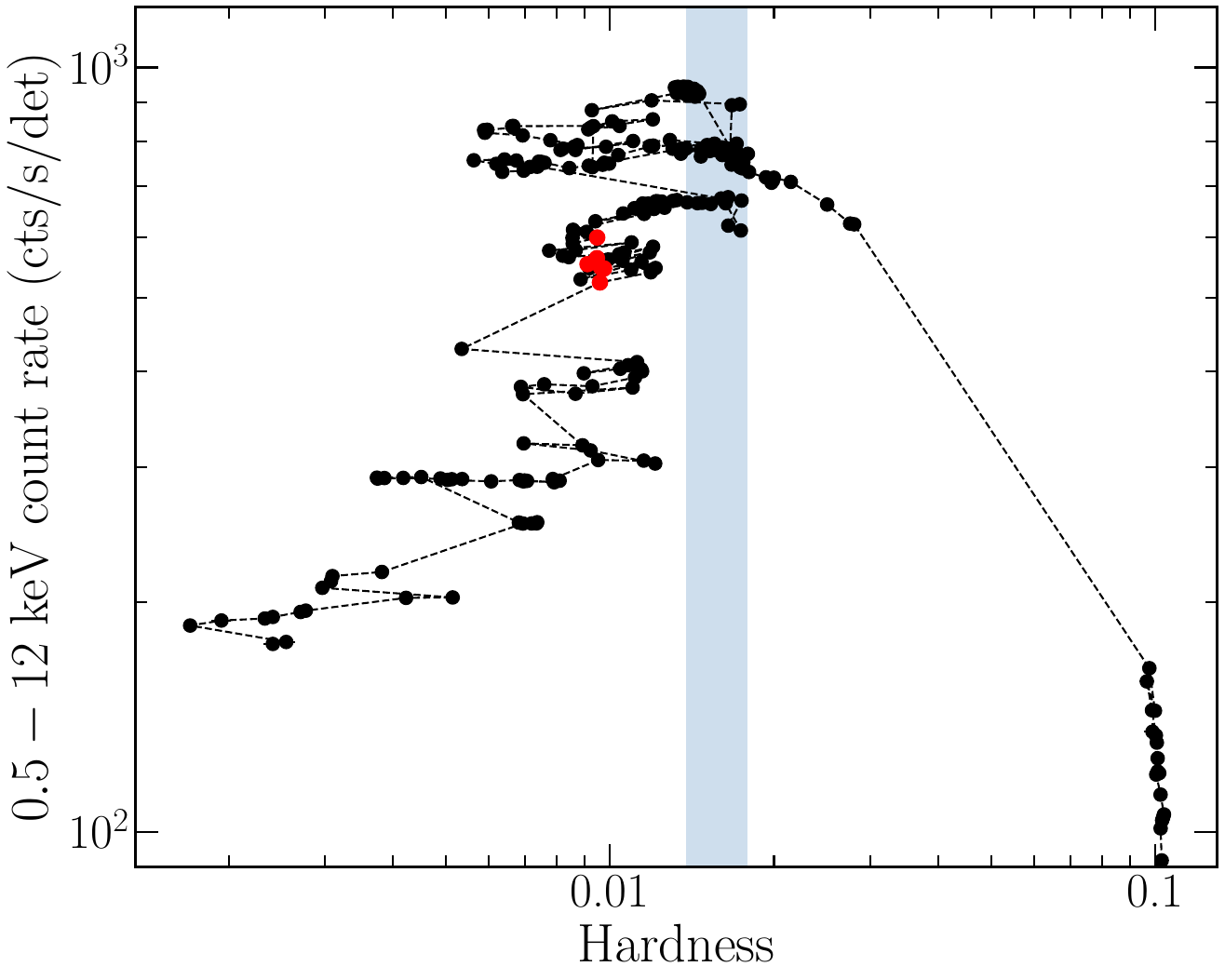}}}
\end{center}
\caption{\nicer\ hardness-intensity diagram (HID) of the main outburst of \target. The hardness is defined as the ratio between the 6--12 keV and 2--3.5 keV count rate. Only data points with a count rate larger than 90 cts s$^{-1}$ det$^{-1}$ are plotted. The red points correspond to the data segments with a type-A QPO. The shaded region marks the range in hardness where a type-B QPO is present \citep[see][]{Belloni2020, Zhang2020a}. Each point corresponds to a separate continuous data segment.}
\label{fig:HID}
\end{figure}

\section{Analysis and results}

\begin{table}
    \centering
    \caption{Log of the \nicer\ observations of \target\ with a type-A QPO. The third column lists the exposure time of the data intervals with the type-A QPO. The fourth column lists the characteristic frequency of the type-A QPO.}
    \label{tab:log}
    \begin{tabular}{lccc}
    \hline\hline
    ObsId & Start time (MJD) & Exposure (s) & QPO freq. (Hz)\\
    \hline
    1200530128 &  58543.65  &  1167  &  7.34 $\pm$ 0.24 \\
    2200530102 &  58545.77  &  1244  &  7.00 $\pm$ 0.31 \\
    2200530103 &  58546.67  &  1112  &  7.10 $\pm$ 0.28 \\
    2200530104 &  58547.38  &  1949  &  7.22 $\pm$ 0.21 \\
    2200530106 &  58549.83  &  790   &  6.77 $\pm$ 0.41 \\
    \hline
    \end{tabular}
\end{table}

\subsection{Timing analysis}
\label{sec:timing}

For each data segment, we extracted an average PDS from intervals of 16 s over the 0.5--10 keV energy band. We used a time resolution of $\sim$1/8192 s so that the Nyquist frequency is $\sim$4096 Hz. The resulting PDS were normalized in units of (rms/mean)$^{2}$~Hz$^{-1}$ \citep{Belloni1990}, and the Poisson noise level estimated from the power between 3000 and 4000 Hz was subtracted. 
Upon examination, we identified a distinct type-A QPO in five of the data segments, resembling the PDS pattern shown in Fig. 4(c) of \citetalias{Zhang2020a}.
In Fig.~\ref{fig:HID}, we show the hardness-intensity diagram (HID) for the data points with a count rate above 90 counts per second per detector. The data segments exhibiting a type-A QPO are highlighted with red points, and the observation log is reported in Tabel \ref{tab:log}. It is apparent that the type-A QPOs occur within a narrow range of hardness values. The hardness value corresponding to the type-A QPO is significantly lower than that of the type-B QPO. In order to obtain the QPO parameters, we fitted the PDS with a model consisting of a sum of Lorentzian functions, as proposed by \citet{Belloni2002}. The best-fitting characteristic frequencies of the QPOs are shown in Table~\ref{tab:log}. The characteristic frequencies are consistent within errors across different observations.

\begin{figure}
\begin{center}
\resizebox{1\columnwidth}{!}{\rotatebox{0}{\includegraphics{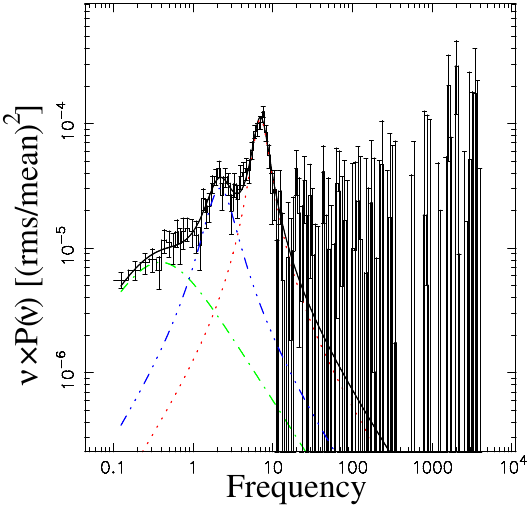}}}
\end{center}
\caption{0.5--10 keV power density spectrum of \target\ obtained by averaging the five data segments listed in Table \ref{tab:log}. The PDS was rms normalized and the contribution due to Poisson noise was subtracted.}
\label{fig:pds}
\end{figure}

Since the QPO peaks appear to be very similar in frequency, we extracted an average PDS by adding all the data segments with the type-A QPO together. This results in a total exposure time of about 6 ksec. The average PDS is shown in Fig.~\ref{fig:pds}. 
The QPO is detected at a significance of $\sim$11~$\sigma$. A fit to the average PDS yields a characteristic frequency of $7.038 \pm 0.126$ Hz, a $Q$ factor of $1.985 \pm 0.248$, and a 0.5--10 keV fractional rms of $0.882 \pm 0.039$ per cent. These parameters exhibit typical characteristics of type-A QPOs.
In addition to the 7-Hz type-A QPO peak, a 2-Hz peak and a weak low-frequency noise component are also seen in the PDS.

\begin{figure}
\begin{center}
\resizebox{1\columnwidth}{!}{\rotatebox{0}{\includegraphics{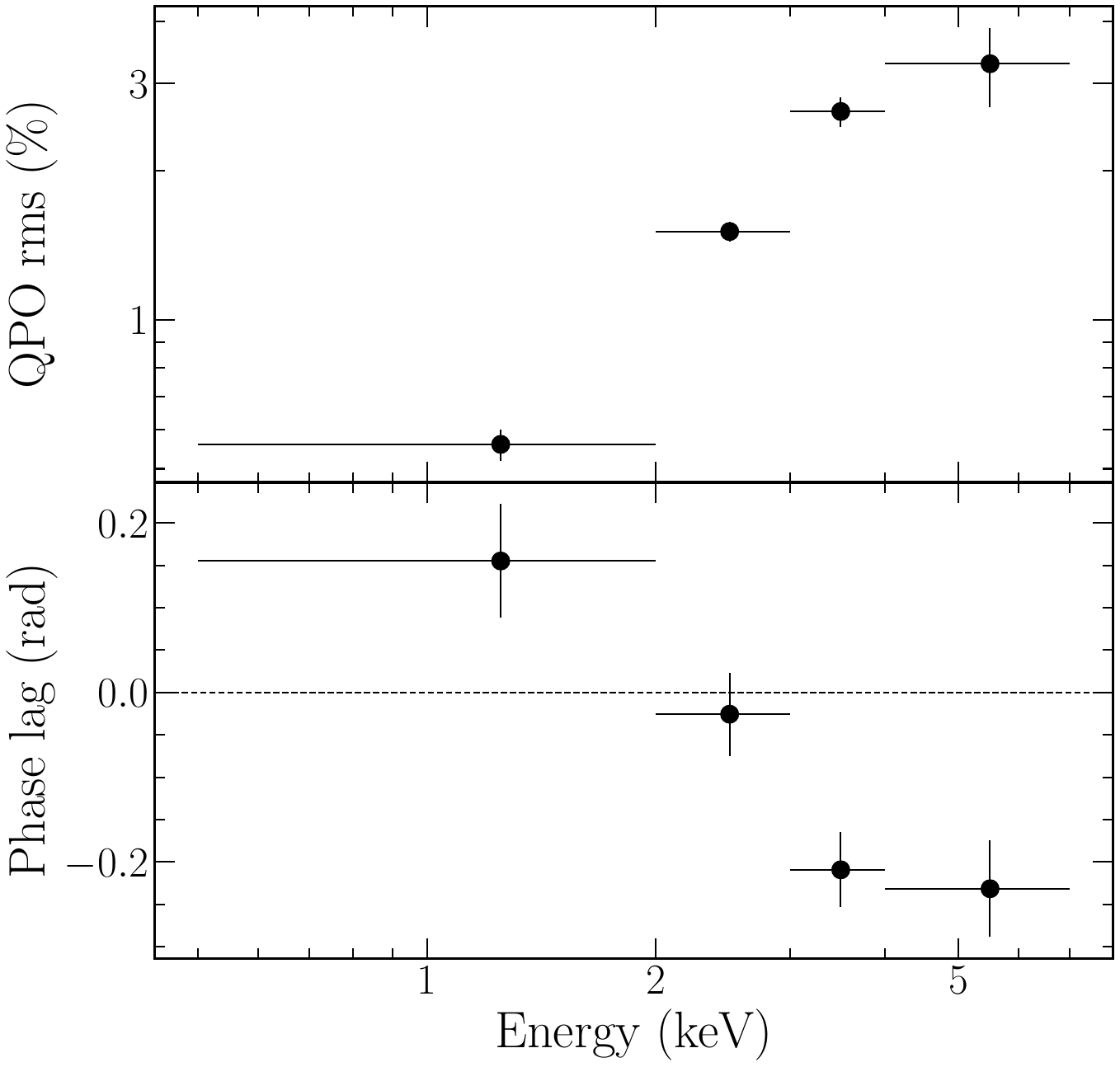}}}
\end{center}
\caption{Fractional rms amplitude (upper panel) and phase lag (lower panel) of the type-A QPO in \target\ as a function of energy. To calculate the phase lags of the QPO we utilized the full energy band (0.5--10 keV) as the reference band.}
\label{fig:rms}
\end{figure}

In order to investigate the energy-dependent properties of the type-A QPO, we extracted PDS for different energy bands following the same procedure as described earlier. No significant QPO signals were detected above 7 keV due to insufficient statistics (the 7.0--10.0 keV count rate is less than 20 cts s$^{-1}$ for all detectors combined). Consequently, we excluded the data above 7 keV from our analysis. We calculated the fractional rms of the type-A QPO in the 0.5--2.0 keV, 2.0--3.0 keV, 3.0--4.0 keV, and 4.0--7.0 keV energy bands. For this, we fitted the same multi-Lorentzian model that we used for the full-band PDS to the PDS in each energy band. In the fitting process, we fixed the centroid frequency and FWHM of the QPO to values determined from the full-band PDS and let the normalizations free to vary.
In the upper panel of Fig.~\ref{fig:rms}, we show the fractional rms of the QPO as a function of energy. It is observed that the fractional rms of the QPO increases with energy from below 1 per cent to $\sim$3 per cent.

To calculate the energy-dependent lags of the type-A QPO, we produced a cross-spectrum for each energy band using the full energy band (0.5--10 keV) as the reference band, following the method described in \citet{Uttley2014}. We computed the phase lags at the QPO frequency by averaging the cross-spectra over a symmetric frequency range centered on the centroid frequency of the QPO with a width of one FWHM. We subtracted the average of the real part of the cross-spectrum over a frequency range dominated by Poisson noise to correct the resulting phase lags for cross-channel talk and partial correlation between the subject bands and the reference band.
In the lower panel of Fig.~\ref{fig:rms}, we show the phase lags of the QPO as a function of energy. We found that the soft photons always lag behind the hard photons.

\begin{figure}
\begin{center}
\resizebox{1\columnwidth}{!}{\rotatebox{0}{\includegraphics{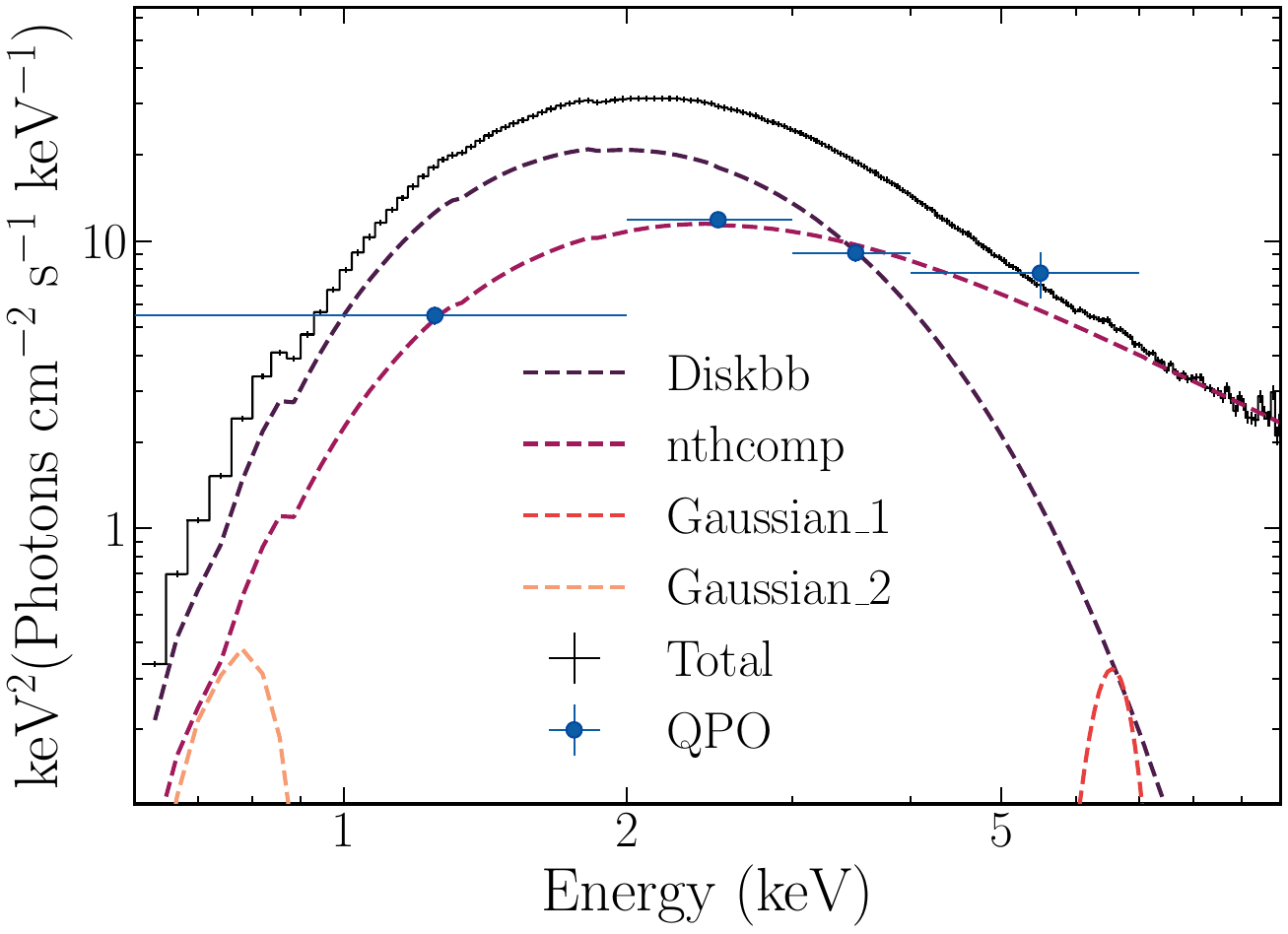}}}
\end{center}
\caption{A representative energy spectrum of \target\ (ObsId: 2200530104) and the type-A QPO spectrum. Individual model components are marked with dotted lines in different colors. The QPO spectrum is scaled by a factor of 400 for better comparison.}
\label{fig:ep}
\end{figure}

\subsection{Spectral analysis}
\label{sec:spectra}

\begin{table*}
\centering
\caption{Best-fitting spectral parameters of the fits to the spectra of \target\ for the data segments with a type-A QPO. The spectra were fitted with the model \textsc{tbnew*(diskbb+nthcomp+gaussian+gaussian)} in the 0.6--10 keV band.}
\label{tab:best_fits}
\begin{tabular}{lcccccc}
\hline\hline
Component   &   Parameter   &   1200530128   &   2200530102   &   2200530103   &   2200530104   &   2200530106 \\
\hline

TBNEW    &  $N_{\rm H}$ ($\times10^{22}~{\rm cm}^{-2}$)    &  $0.94 \pm 0.01$  &  $0.93 \pm 0.01$  &  $0.93 \pm 0.01$  
                                                           &  $0.94 \pm 0.01$  &  $0.94 \pm 0.01$\\      
                                                      
DISKBB   &  $kT_{\rm in}$ (keV)    &  $0.62 \pm 0.01$  &  $0.61 \pm 0.01$  &  $0.61 \pm 0.01$  
                                   &  $0.60 \pm 0.01$  &  $0.60 \pm 0.01$\\
                                
         &  Norm                   &  $35582 \pm 386$  &  $35280 \pm 403$  &  $36565 \pm 424$  
                                   &  $35779 \pm 391$  &  $35644 \pm 479$\\
                                
NTHCOMP  &  $\Gamma$               &  $3.49 \pm 0.04$  &  $3.48 \pm 0.04$  &  $3.45 \pm 0.04$  
                                   &  $3.49 \pm 0.03$  &  $3.43 \pm 0.05$\\
                                
         &  Norm                   &  $11.73 \pm 0.48$ &  $11.13 \pm 0.46$ &  $9.95 \pm 0.46$  
                                   &  $11.46 \pm 0.37$ &  $10.24 \pm 0.55$\\
                                
GAUSSIAN\_1 &  $E_{\rm line}$ (keV)   &  $6.50 \pm 0.04$  &  $6.63 \pm 0.09$  &  $6.51 \pm 0.10$  
                                   &  $6.52 \pm 0.05$  &  $6.42 \pm 0.15$\\
                                
         &  $\sigma$ (keV)         &  $0.23 \pm 0.07$  &  $0.45 \pm 0.10$  &  $0.48 \pm 0.10$  
                                   &  $0.34 \pm 0.06$  &  $0.58 \pm 0.15$\\
                                
    &  Norm (ph~cm$^{-2}$~s$^{-1}$)&  $0.005 \pm 0.001$ & $0.007 \pm 0.002$ & $0.008 \pm 0.002$ 
                                   &  $0.007 \pm 0.001$ & $0.009 \pm 0.003$\\
         
GAUSSIAN\_2 &  $E_{\rm line}$ (keV)   &  $0.71 \pm 0.01$  &  $0.72 \pm 0.01$  &  $0.72 \pm 0.01$  
                                   &  $0.71 \pm 0.01$  &  $0.72 \pm 0.01$   \\
                                
         &  $\sigma$ (keV)         &  $0.07 \pm 0.01$  &  $0.07 \pm 0.01$  &  $0.07 \pm 0.01$  
                                   &  $0.07 \pm 0.01$  &  $0.07 \pm 0.01$\\
                                
    &  Norm (ph~cm$^{-2}$~s$^{-1}$)&  $2.83 \pm 0.37$  &  $2.67 \pm 0.33$  &  $2.65 \pm 0.34$  
                                   &  $2.96 \pm 0.35$  &  $2.63 \pm 0.38$\\
\hline
   & $F_{\rm diskbb}^{*}$ (erg cm$^{-2}$ s$^{-1}$)   &  $8.21 \times 10^{-8}$  &  $7.68 \times 10^{-8}$  &  $7.88 \times 10^{-8}$
                                   &  $7.34 \times 10^{-8}$  &  $7.21 \times 10^{-8}$\\

   & $F_{\rm nthcomp}^{*}$ (erg cm$^{-2}$ s$^{-1}$)  &  $4.54 \times 10^{-8}$  &  $4.27 \times 10^{-8}$  &  $3.84 \times 10^{-8}$
                                   &  $4.34 \times 10^{-8}$  &  $3.92 \times 10^{-8}$\\

   & $F_{\rm total}^{*}$ (erg cm$^{-2}$ s$^{-1}$)    &  $1.31 \times 10^{-7}$  &  $1.23 \times 10^{-7}$  &  $1.20 \times 10^{-7}$
                                   &  $1.20 \times 10^{-7}$  &  $1.14 \times 10^{-7}$\\

\hline
         &  $\chi^{2}/{\rm dof}$   &    132.18/153     &    137.59/153     &    135.25/152     &   151.16/157   &   132.83/148\\ 
\hline
\multicolumn{7}{l}{$^*$ 0.6--10 keV unabsorbed flux.}
\end{tabular}
\end{table*}

For each data segment with a type-A QPO, we extracted the total and background spectra using the \nicer\ Background Estimator Tool \texttt{nibackgen3C50}\footnote{\url{https://heasarc.gsfc.nasa.gov/docs/nicer/tools/nicer_bkg_est_tools.html}}. We binned the spectra in accordance with the optimal binning algorithm proposed by \citet{Kaastra2016} with a minimum of 30 counts per energy bin. We added a systematic error of 1 per cent to energies below 3 keV to reduce the effects of calibration uncertainties at low energies. The detector redistribution matrix file and the auxiliary response file were generated with the tasks \texttt{nicerrmf} and \texttt{nicerarf}, respectively. We fitted the spectra in the 0.6--10 keV band using \texttt{xspec} version 12.13.0.

We conducted spectral fitting using a model consisting of a multi-temperature blackbody component (\textsc{diskbb}: \citealt{Mitsuda1984}) and a thermally Comptonised continuum (\textsc{nthcomp}: \citealt{Zdziarski1996, Zycki1999}), plus an emission line at around 6.5 keV representing the iron $K_{\alpha}$ line~\citep{1989MNRAS.238..729F}. All of these components were modified by interstellar absorption. 
The seed photon temperature of the Comptonised component, $kT_{\rm bb}$, was linked to the inner disc temperature of the disc component, $kT_{\rm in}$. As the high energy cutoff of the Comptonised component exceeds the \nicer\ energy range, we fixed the electron temperature at 250 keV. The absorption along the line of sight was modeled using \textsc{tbnew}, which is an improved version of the X-ray absorption model \textsc{tbabs}\footnote{\url{https://pulsar.sternwarte.uni-erlangen.de/wilms/research/tbabs/}}. The oxygen and iron absorption abundances were fixed at the solar value. 
Since extra residuals were still clearly seen at around 0.7 keV, we added another \textsc{gaussian} to account for these residuals. The best-fitting spectral parameters obtained from the fitting are listed in Table~\ref{tab:best_fits}. A representative spectrum is shown in Fig.~\ref{fig:ep}.

The best-fitting spectral parameters are similar for all data segments. The spectra are dominated by a disc component with an inner disc temperature of around 0.6 keV, suggesting that the source was in a soft state during this period. The Comptonised component is very steep with a power-law photon index of around 3.5.

In order to compare the total time-averaged energy spectrum to that of the fast variability, we converted the absolute rms amplitude of the type-A QPO to the same units as the total spectrum by considering the response of the instrument and constructed a QPO spectrum. The resulting QPO spectrum is shown in Fig.~\ref{fig:ep} with blue points. It is apparent that the shape of the QPO spectrum is similar to that of the Comptonised component, indicating that the Comptonised region is driving the variability.

\begin{figure*}
\begin{center}
\resizebox{2\columnwidth}{!}{\rotatebox{0}{\includegraphics{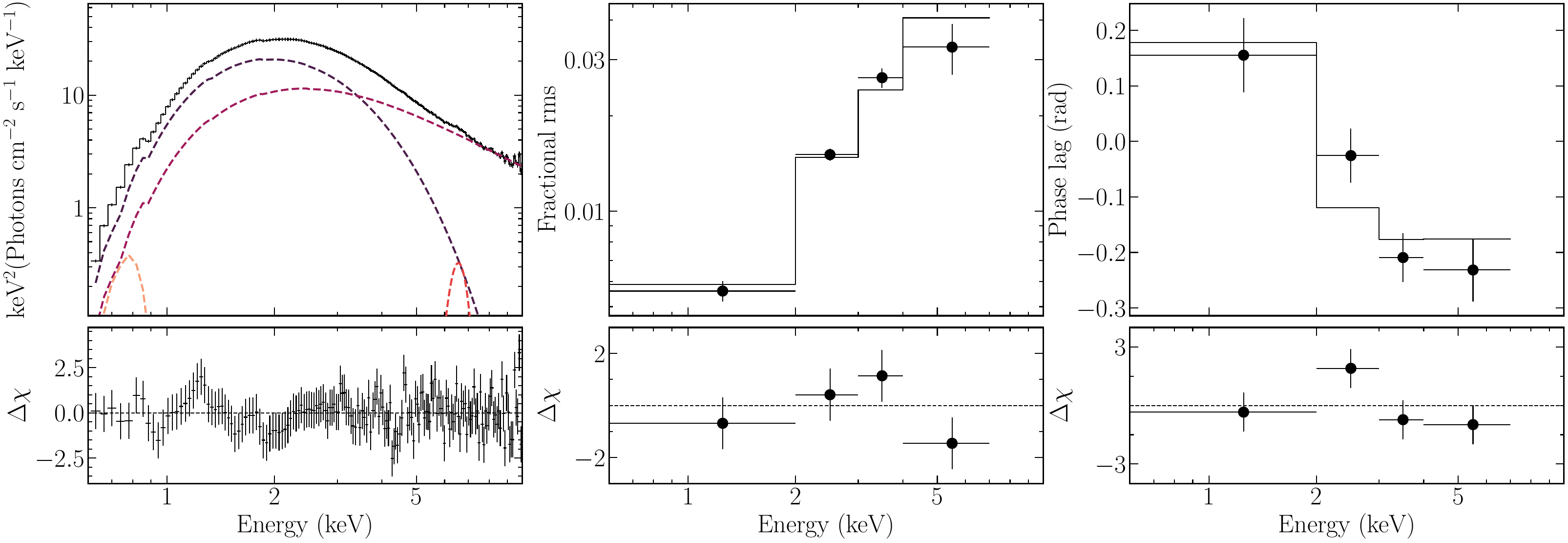}}}
\end{center}
\caption{Time-averaged energy spectrum (left panel) of \target\ obtained from ObsID 2200530104, and the fractional rms amplitude (middle panel) and phase-lag (right panel) spectra of the type-A QPO. In the middle and right panels, the solid lines represent the best-fitting model with \textsc{vkompthdk}. Bottom panels: residuals.}
\label{fig:vkompth}
\end{figure*}

\subsection{Joint fitting of the time-averaged, rms and phase-lag spectra}

In this section, we fitted jointly the time-averaged energy spectrum of \target, and the rms and phase-lag spectra of the type-A QPO, using the single-component Comptonisation model \textsc{vkompthdk} developed by \citet{Bellavita2022}.

The \textsc{vkompthdk} model\footnote{We note that the steady state spectrum of \textsc{vkompthdk} is the same as that of \textsc{nthcomp}.} considers that the source of the seed photons is a geometrically thin and optically thick accretion disc \citep{Shakura1973} with a temperature $kT_{\rm s}$ at the inner edge. These seed photons are then subjected to inverse-Compton scattering within a spherically symmetric corona, characterized by a size $L$ and temperature $kT_{\rm e}$. The corona is maintained in thermal equilibrium through heating from an external source.
Additionally, the model incorporates a feedback process, where a portion of the photons scattered into the corona are redirected back onto the accretion disc. The feedback fraction, $0 \leq \eta \leq 1$, in the model denotes the fraction of the disc flux that is due to feedback from the corona \citep[see][]{Karpouzas2020}. The feedback fraction, $\eta$, is related to the intrinsic feedback fraction, $\eta_{\rm int}$, which denotes the fraction of the photons emitted by the corona that returns to the accretion disc. 
This model treats the QPO as small oscillations of the spectrum around the time-averaged one, assuming that fluctuations of the spectrum are caused by perturbations in the electron temperature, $kT_{\rm e}$, via feedback, of the seed photon source temperature, $kT_{\rm s}$, as a result of an oscillating external heating rate, $\delta \dot{H}_{\rm{ext}}$ \citep[see][for a more detailed description of the model]{Bellavita2022}.
The main parameters of the model are the seed photon source temperature, $kT_{\rm s}$, the electron temperature, $kT_{\rm e}$, the power-law photon index, $\Gamma$, the size of the corona, $L$, the feedback fraction, $\eta$, and the variation of the external heating rate, $\delta \dot{H}_{\mathrm{ext}}$.

In the joint fit, we used the same model described in Section \ref{sec:spectra} to fit the time-averaged spectrum of \target. The \textsc{vkompthdk} model is loaded as an external component to fit the rms and lag spectra of the QPO.
We initially tried to link the seed photon source temperature of \textsc{vkompthdk}, $kT_{\rm s}$, to the inner disc temperature of \textsc{diskbb}, $kT_{\rm in}$. Additionally, the electron temperature and power-law photon index of \textsc{vkompthdk} were linked to the corresponding parameters of \textsc{nthcomp}.
Following \citet{Bellavita2022}, we introduced a dilution correction to the fractional rms amplitude computed in the model to take into account the fraction of the non-variable emission.  
The initial fit (Case 1) yields a $\chi^{2}/{\rm dof}$ value of 166.0/161 ($\chi^{2}=14.8$ for the 8 bins of the rms and lag spectra). We note that the $\chi^{2}$ and dof are dominated by the time-averaged spectrum. The main residuals come from the fit to the rms and lag spectra.
Allowing $kT_{\rm s}$ and $kT_{\rm in}$ to vary independently (Case 2) improves the fit to the rms and lag spectra significantly with $\chi^{2}=9.3$ for the 8 bins of the rms and lag spectra.
We obtained a corona size of $\sim$2300 km ($\sim$150 $R_{\rm g}$ for a 10-$M_{\sun}$ black hole), with an intrinsic feedback fraction of $\eta_{\rm int} \approx 21$ per cent. The seed photon source temperature of the corona is $kT_{\rm s} = 0.37 \pm 0.07$ keV, lower than the temperature of the inner disc, $kT_{\rm in} = 0.60 \pm 0.01$ keV.
The best-fitting parameters of both cases are presented in Table \ref{tab:joint_fits} and the best-fitting \textsc{vkompthdk} model of Case 2 is shown in Fig. \ref{fig:vkompth}. 
From this figure, it is apparent that the \textsc{vkompthdk} model describes the rms and lag spectra of the type-A QPO well.

\section{Discussion and Conclusions}

\begin{table}
\centering
\caption{Best-fitting parameters of the joint fit to the time-averaged energy spectrum of \target, and the rms and lag spectra of the type-A QPO, using the single-component Comptonisation model \textsc{vkompthdk}.}
\label{tab:joint_fits}
\begin{tabular}{lccc}
\hline\hline
Component   &   Parameter   &   Case 1   & Case 2\\
\hline

TBNEW    &  $N_{\rm H}$ ($\times10^{22}~{\rm cm}^{-2}$)    &  $0.94 \pm 0.01$  &  $0.94 \pm 0.01$\\      
                                                      
DISKBB   &  $kT_{\rm in}$ (keV)    &  $0.60 \pm 0.01$  &  $0.60 \pm 0.01$\\
                                
         &  Norm                   &  $35828 \pm 393$  &  $35812 \pm 392$\\
                                
NTHCOMP  &  $\Gamma$               &  $3.49 \pm 0.03$  &  $3.49 \pm 0.03$\\
                                
         &  Norm                   &  $11.44 \pm 0.38$ &  $11.44 \pm 0.38$\\
                                
GAUSSIAN\_1 &  $E_{\rm line}$ (keV)   &  $6.52 \pm 0.04$  &  $6.52 \pm 0.04$\\
                                
         &  $\sigma$ (keV)         &  $0.34 \pm 0.06$  &  $0.34 \pm 0.06$\\
                                
         &  Norm                   &  $0.007 \pm 0.001$ & $0.007 \pm 0.001$\\
         
GAUSSIAN\_2 &  $E_{\rm line}$ (keV)   &  $0.71 \pm 0.01$  &  $0.71 \pm 0.01$\\
                                
         &  $\sigma$ (keV)         &  $0.07 \pm 0.01$  &  $0.07 \pm 0.01$\\
                                
         &  Norm                   &  $2.97 \pm 0.35$  &  $2.96 \pm 0.35$\\

VKOMPTHDK &    $kT_{\rm s}$ (keV)  &    $= kT_{\rm in}$   &  $0.37 \pm 0.07$          \\
          &    $L$ (km)            &  $1959 \pm 422$     &  $2304 \pm 524$\\
          &    $\eta$              &  $0.47 \pm 0.04$    &  $0.60 \pm 0.07$\\
          &    $\eta_{\rm int}$    &  $0.17 \pm 0.01$    &  $0.21 \pm 0.01$\\
          &    $\delta \dot{H}_{\rm{ext}}$  &  $0.09 \pm 0.01$  & $0.05 \pm 0.01$\\ 
\hline
         &  $\chi^{2}/{\rm dof}$   &    166.0/161    &  160.4/160 \\ 
\hline
\end{tabular}
\end{table}

We have reported a comprehensive spectral-timing analysis of a type-A QPO observed in the HSS of \target.
The QPO is characterized by a broad peak ($Q \approx 2$) at a frequency of approximately 7 Hz, with an integrated fractional rms amplitude of 0.9 per cent in the 0.5--10 keV band.
The large effective area of \nicer, along with the high flux of the source and the relatively long accumulative exposure time, enables us to construct the first rms and lag spectra for a typical type-A QPO and perform the first study of a type-A QPO with information in the 0.5--2.0 keV.
Below we discuss the origin of the QPO rms and lag, and the geometry of the Comptonised region.

\subsection{The rms and lag spectra of the type-A QPO}

The fractional rms amplitudes of both the type-C and type-B QPOs of \target\ display an initial rise with increasing energy, after which they remain more or less constant with a typical rms of 10--20 per cent. \citep{Belloni2020, Alabarta2022, Liu2022}. Similar QPO rms spectra have been observed in the type-C QPOs of many other BHXBs \citep[e.g.,][]{Zhang2017, Huang2018, Mendez2022, Ma2023}. 
At energies above 10 keV, the contribution of the disc component to the X-ray spectrum becomes negligible; the hard X-ray emission is likely due to the Comptonisation of the soft disc photons by the hot electrons in a corona.
The large QPO rms amplitude observed at these high energies suggests that it is the corona that drives the variability rather than the disc.  
Frequency-resolved spectroscopy conducted by \citet{Axelsson2016} reveals that the rms spectrum of the type-C QPO does not show signs of the disc component, even when the disc is prominent in the time-averaged spectrum.
Despite the type-A QPO being significantly weaker in comparison to the type-C and type-B QPOs in \target, we observe that its fractional rms amplitude also increases with energy. Additionally, we find that the shape of the QPO spectrum closely resembles that of the Comptonised component. These findings imply that the variability associated with the type-A QPO is also generated within the corona.

The phase-lag characteristics of the type-A QPO differ significantly from those of the type-C and type-B QPOs in the same source.
\citet{Alabarta2022} found that the type-C QPOs of \target\ always exhibit hard lags, indicating a delay of the hard photons compared to the soft photons. The hard lags can be naturally explained by the Comptonisation process.
On the other hand, \citet{Belloni2020} discovered that for the type-B QPO of \target, photons at all energies lag behind those in the 2--2.5 keV band.
However, in the case of the type-A QPO of \target, we find that the soft photons consistently lag behind the hard photons up to $\sim$7 keV.
The soft QPO lags can be explained by the feedback process, where a portion of the high-energy photons emitted by the corona are redirected back to the accretion disc, and subsequently re-emitted at lower energies \citep{Lee2001}.

\subsection{The geometry of the Comptonised region}

The energy-dependent fractional rms amplitude and phase lags of the type-B QPO in \target\ can be described by a time-dependent Comptonisation model consisting of two physically-connected regions: a large ($\sim$12000 km) jet-like region and a small ($\sim$150 km) horizontally extended region \citep{Garcia2021, Bellavita2022}. Similarly, the rms and phase-lag spectra of the type-B QPO in MAXI J1820+070 can also be well-fitted with a model consisting of a small ($\sim$20 km) and a large ($\sim$10$^{4}$ km) coronae that are physically connected \citep{Ma2023b}.
In a different source, MAXI J1535--571, the radiative properties of the type-B QPO can be explained by a single jet-like Comptonised region with a size of $\sim$6500 km \citep{Zhang2023}.
By jointly fitting the time-averaged spectrum of \target, as well as the rms and phase-lag spectra of the type-A QPO with the single-component Comptonisation model \textsc{vkompthdk}, we obtain a corona size of $\sim$2300 km, which is smaller than the jet-like region associated with the type-B QPO.
The seed photon source temperature of the corona, $kT_{\rm s}$, was found to be lower than the temperature of the inner disc, $kT_{\rm in}$. This can be explained by considering the Comptonised region as a jet-like structure that is illuminated by the outer parts of the disc. In this scenario, the outer parts of the disc supply relatively cold seed photons, while the corresponding Comptonised photons with a longer light-crossing time primarily contribute to the observed QPO lags; on the other hand, the inner parts of the disc provide the seed photons and the Comptonised photons mainly contribute to the steady-state spectrum \citep[see][]{Peirano2023}.

It is commonly observed that the type-A QPO tends to occur slightly later than the type-B QPO \citep[e.g.,][]{Casella2005, Motta2011}.  In light of this, it is possible that the Comptonised region associated with the type-A QPO evolves from that of the type-B QPO.
If this is indeed the case, our result suggests that the jet-like corona is shrinking in size from the SIMS to the HSS.
Radio observations of \target\ have revealed that the radio flux drops significantly from several tens of mJy to undetectable levels from the SIMS to the HSS \citep{Carotenuto2021}. The decrease in radio flux could potentially be attributed to the shrinking of the vertically extended jet-like corona.
Furthermore, we determined an intrinsic feedback fraction of around 0.21, indicating that approximately 21 per cent of the Comptonised photons impinge back onto the accretion disc. This suggests that the corona partially covers the accretion disc.

It is important to note that the Comptonisation model \textsc{vkompthdk} used in this work is based on the assumption that the corona is spherically symmetric and homogeneous with a constant temperature and optical depth. However, it is likely that the actual geometry of the corona is more complex \citep{Kawamura2022}. Consequently, the corona size values measured from the model should be regarded as a characteristic size of the corona rather than the precise radius of a spherical corona \citep[see][]{Mendez2022}.

\section*{Acknowledgements}

This work was supported by the National Natural Science Foundation of China (NSFC) under grants 12203052 and U2038104.
MM acknowledges the research programme Athena with project number 184.034.002, which is (partly) financed by the Dutch Research Council (NWO). 
MM and FG thank the Team Meeting at the International Space Science Institute (Bern) for fruitful discussions. This research was supported by the International Space Science Institute (ISSI) in Bern, through ISSI International Team project \#486.
FG is a researcher of CONICET and acknowledges support from PIBAA 1275 and PIP 0113 (CONICET).
YZ acknowledges support from the China Scholarship Council (CSC 201906100030).
D.A. acknowledges support from the Royal Society. 
This work was supported by NASA through the \nicer\ mission and the Astrophysics Explorers Program, and made use of data and software provided by the High  Energy Astrophysics Science Archive Research Center (HEASARC).

\section*{Data Availability}

The data underlying this article are available in the HEASARC database.

\bsp	
\label{lastpage}
\end{document}